\documentclass[aps,showpacs,twocolumn,prl]{revtex4}

\usepackage{amsmath}
\usepackage{amsthm}
\usepackage{amsfonts}
\usepackage{amssymb}
\usepackage{graphicx}

\begin{document}
\title{Mechanism of delayed double ionization in a strong laser field}

\author{F. Mauger$^1$, A. Kamor$^2$, C. Chandre$^1$, T. Uzer$^2$}
\affiliation{$^1$ Centre de Physique Th\'eorique, CNRS -- Aix-Marseille Universit\'e, Campus de Luminy, case 907, F-13288 Marseille cedex 09, France \\  $^2$ School of Physics, Georgia Institute of Technology, Atlanta, GA 30332-0430, USA}

\begin{abstract}
When intense laser pulses release correlated electrons, the time delay between the ionizations may last more than one laser cycle. We show that this ``Recollision-Excitation with Subsequent Ionization'' pathway originates from the inner electron being promoted to a sticky region by a recollision where it is trapped for a long time before ionizing. We identify the mechanism which regulates this region, and predict oscillations in the double ionization yield with laser intensity.
\end{abstract}
\maketitle

Atoms in strong laser pulses lose their electrons through two ionization channels~\cite{Beck08}: A sequential one (SDI) and its non-sequential counterpart (NSDI). The SDI mechanism consists of successive and independent removals of the electrons. The recollision (or three-step) scenario~\cite{Cork93,Scha93} in which a pre-ionized electron returns to the ion core to dislodge a bound electron is the characteristic mechanism for NSDI. It turns out that there is a rich variety of pathways among NSDI processes, also. At first sight, these pathways can be distinguished by the time it takes the second electron to ionize. The common variant involves little, if any, delay between the recollision and ionization. However, this so-called ``direct impact ionization''~\cite{Rude04} is often accompanied by an alternative (and less straightforward) road to NSDI called Recollision Excitation with Subsequent Ionization (or RESI for short~\cite{Haan08,Haan10,Shaa10,Baie08,Feue01,Rude04,DeJe04_1,DeJe04_2,Figu11}). Recent experiments have shown that RESI can be the dominant channel for very short pulses~\cite{John11}. The mechanism for RESI is often attributed to the recollision which excites the parent ion, later ionized by the laser field with a delay sometimes lasting longer than one laser cycle after the recollision and thereby imitating an uncorrelated ionization process. 

\begin{figure}
	\centering
		\includegraphics[width = \linewidth]{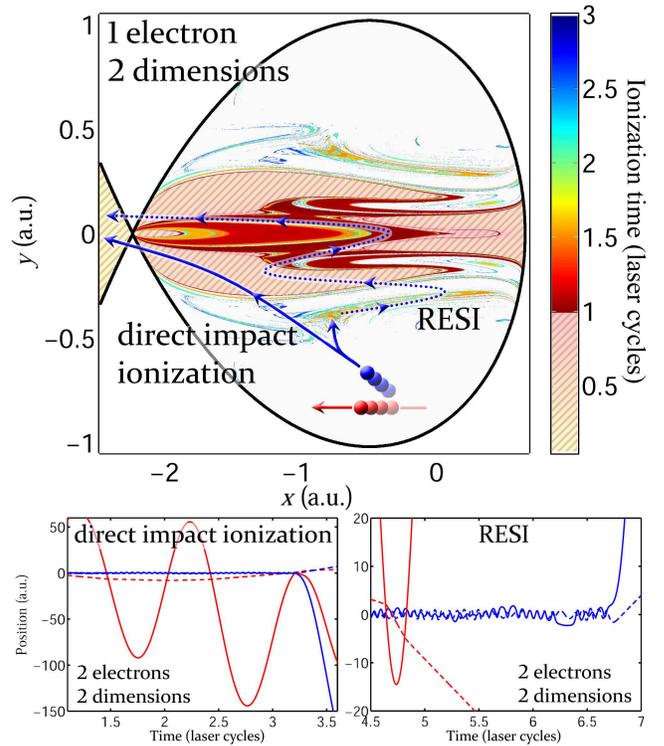}
	\caption{\label{fig:RESI_mechanism}
	Upper panel: Ionization time for Hamiltonian~(\ref{eq:He+}) with two dimensions for $I=3\times10^{15}\ {\rm W}\cdot{\rm cm}^{-2}$, $\phi_{0}=\pi/2$ and $780$~nm. Initial conditions are chosen at the energy of the Stark saddle~\cite{ChaosAtomPhys} with $p_{x}\leq0$ and $p_{y}=0$. After a recollision (full line arrows), the inner electron can be directly ionized or promoted to an excited state that is ionized with a delay (dotted arrow).
	Lower panels: Typical two-electron trajectories for direct impact ionization (left) and RESI (right). Continuous (dashed) curves denote $x$ ($y$) coordinates.
	}
\end{figure}

Here we report results showing that RESI is a purely inner electron process and that the recollision, while helpful for increasing its probability, merely enhances what is already there, namely the ionization process due to extensive state mixing. The reason why the mechanism responsible for RESI remains an open question is due to the spectator role often erroneously assigned to the inner electron. Far from waiting passively for the pre-ionized electron to recollide, the inner electron interacts with the strong field also, and a true nonsequential description of the process reflects the excitation of both electrons. Our view, to be detailed below, is summarized by the cartoon in Fig.~\ref{fig:RESI_mechanism}: Hitting the pool of available initial states, a recollision may send the inner electron into the ionization continuum directly, thereby causing two electrons to emerge with very little time delay (direct impact ionization) or the recollision may drive the bound electron towards the parts of phase space which funnel electrons into the ionization continuum -- which, given sufficient time, the electrons may have reached without the help of a recollision.

This common-sense distinction between the direct impact ionization and RESI emerges from our classical calculations~\cite{Maug09} which allow both electrons to interact with the field and are, therefore, fully nonsequential. Evidence for the veracity of the proposed mechanism comes from two observations: The ionization delays are in harmony with experimental observations, and our treatment confirms that the second ionization in RESI happens at the extrema of the laser field~\cite{Webe00_2,Mosh00,Feue01,Rude04,Ruiz08}.

Classical ensemble methods~\cite{Panf01,Ho05_1} have been remarkably successful in identifying direct impact ionization as well as RESI pathways that lead to double ionization  and reproduce the experimental and computational observations closely. Therefore we consider a two active electron atom with soft-Coulomb potentials subjected to an intense and short linearly polarized laser pulse in the dipole approximation~\cite{Panf01,Ho05_1,Java88,Haan94,Maug09}. The Hamiltonian is:
\begin{eqnarray}
   && {\mathcal H} \left( {\bf x}_1, {\bf x}_2, {\bf p}_{1}, {\bf p}_{2}, t \right) = 
      \frac{\left\|{\bf p}_{1}\right\|^{2}}{2} + \frac{\left\|{\bf p}_{2}\right\|^{2}}{2}
      - \frac{2}{\sqrt{\left\|{\bf x}_{1}\right\|^{2} + a^{2}}} \nonumber \\
      && \qquad - \frac{2}{\sqrt{\left\|{\bf x}_{2}\right\|^{2} + a^{2}}} +
      \frac{1}{\sqrt{\left\|{\bf x}_{1}- {\bf x}_{2}\right\|^{2} + b^{2}}} \nonumber \\
      && \qquad + \left( {\bf x}_{1} + {\bf x}_{2}\right) \cdot {\bf e}_{x} E_0 f(t) \sin \omega t, \label{eq:Ham2e}
\end{eqnarray}
where ${\bf x}_i$ is the position vector of the $i$th electron in $d$ dimensions and ${\bf p}_i$ is its canonically conjugate momentum. The linearly polarized (along the $x$-direction with unit vector ${\bf e}_{x}$) laser field is characterized by its amplitude $E_{0}$ and has a wavelength of 780~nm or 460~nm ($\omega=0.0584$ or $0.1~\mbox{a.u.}$ respectively) with a shape $f\left(t\right)$ consisting of two-cycle linear ramp-up and six laser cycle constant plateau. The constants $a$ and $b$ are the electron-nucleus and electron-electron softening parameters respectively. For the computations considered in this paper, we choose $a=b=1$~\cite{Maug09, Panf01, Ho05_1, Java88, Haan94}, even though qualitatively similar results are observed with different softening parameters.

To begin with, we investigate the dynamics when both electrons are confined to a single dimension along the axis of polarization~\cite{note_1}, i.e., $d=1$. The motivation for studying the one-dimensional model is twofold: First, the laser field drives the dynamics along the polarization axis by which ionization is naturally expected to happen. Second, we find that this simplified dynamics forms the skeleton of higher-dimensional dynamics, which, however, differs from the single-dimensional dynamics in significant ways, to be specified below. It should be noted that, three-dimensional calculations give similar results to two-dimensional ones due to the cylindrical symmetry of the problem around the polarization axis.

\begin{figure}
	\centering
		\includegraphics[width = \linewidth]{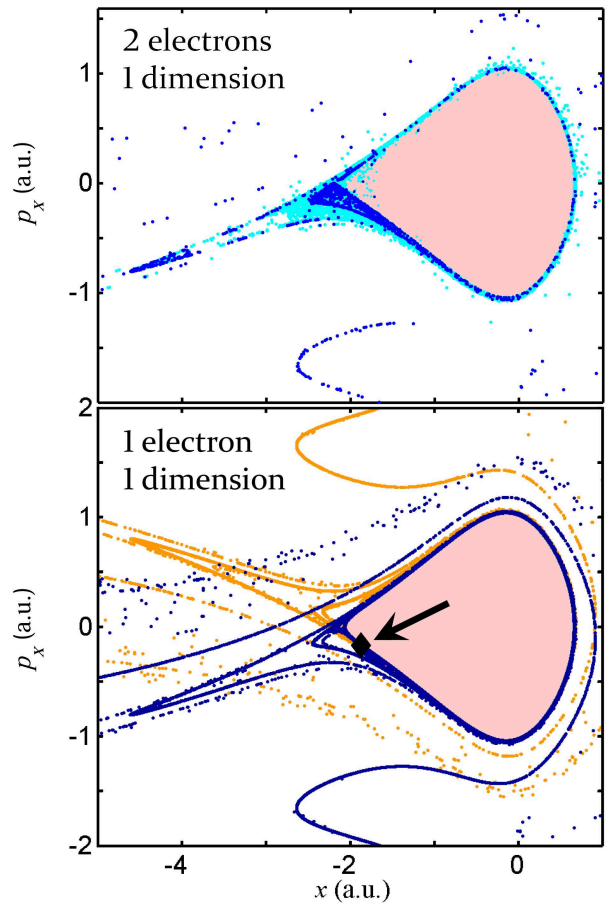}
	\caption{\label{fig:PSOS_RESI1}
	Stroboscopic plots at the maxima of the laser field, for $I=3\times 10^{15}~{\rm W}\cdot{\rm cm}^{-2}$ and $780$~nm during the plateau of the laser. Upper panel: detected RESI trajectories, after the last recollision, for Hamiltonian~(\ref{eq:Ham2e}). Initial conditions are chosen as in Refs.~\cite{Maug09,Maug10}. For each RESI, the first point on the section is plotted in light blue while the following ones are in dark blue. Lower panel: stable (light orange dots) and unstable (dark blue dots) manifolds of the periodic orbit $\mathcal{O}_{12}$ of Hamiltonian~(\ref{eq:He+}). The position of $\mathcal{O}_{12}$ is indicated with a black diamond (see the arrow). The pink area in the panels represents the part of phase space from which the inner electron does not ionize.
	}
\end{figure}

RESI events are described by two-electron trajectories for which, after a recollision, one electron remains bound to the nucleus for a long time before ionizing without any additional recollision, i.e., without the other electron ever returning to the nucleus. We use an energy criterion to designate an electron as ionized or not~\cite{Maug09,Maug10}. A recollision is said to have occurred whenever the distance between the two electrons is smaller than some threshold. We consider the NSDI trajectories which double-ionize after a last recollision. Among them, we discriminate RESIs by a time delay of at least two laser cycles between the last recollision and the ionization of the remaining ion. Qualitatively similar results are obtained with different (shorter or longer) time delays. In Fig.~\ref{fig:PSOS_RESI1} (upper panel), we display stroboscopic coordinate-momentum plots (taken at the maxima of the field) of detected RESIs. In what follows, we explain the rationale for these swirling patterns and connect them to full-dimensional calculations.

RESIs are best understood using a single electron model: A recollision has excited the inner electron while the recolliding electron remains ionized~\cite{Maug09}, which allows us to neglect the electron-electron interaction in Hamiltonian~(\ref{eq:Ham2e}). The resulting reduced-dimensional Hamiltonian reads
\begin{eqnarray} \label{eq:He+}
   &&{\mathcal H} \left( {\bf x}, {\bf p}, t \right) = \frac{\left\|{\bf p}\right\|^{2}}{2} - 
      \frac{2}{\sqrt{\left\|{\bf x}\right\|^{2} + a^{2}}} \nonumber \\
      &&\quad + {\bf x} \cdot {\bf e}_{x} E_{0}\sin(\omega t+\phi_0),
\end{eqnarray}
where $\phi_0$ denotes the phase at which the last recollision occurs.

The dynamics given by Hamiltonian~(\ref{eq:He+}) turns out to be the key to the patterns formed by the RESI trajectories. Previous studies on Hamiltonian~(\ref{eq:He+})~\cite{Maug09,Maug10} with one spatial dimension have identified two qualitatively different kinds of dynamics for the electron driven by the field: In the competition between Coulomb attraction to the nucleus and the laser excitation, either the latter prevails, and the electron is quickly ionized; or the Coulomb attraction manages to maintain the electron trapped near the core~\cite{Hu97}. Two areas in phase space emerge from this distinction: A bound region, close to the nucleus (pink area in Fig.~\ref{fig:PSOS_RESI1}), where the electron is trapped by the nucleus and cannot ionize; and an unbound region, further away (white area in the same panels), where the electron is quickly ionized by the laser. 

A more detailed study shows that the behavior in the unbound region is more complex than anticipated. This is readily apparent in the patterns seen in the unbound region in Fig.~\ref{fig:PSOS_RESI1}. A thin transition layer in the unbound region, and located in the area where the Coulomb attraction and the laser excitation compete equally, is responsible for RESI. In practice, we show that this transition region is organized by the main resonances between the free field dynamics~[$E_{0}=0$ in Hamiltonian~(\ref{eq:He+})] and the laser (the electron revolves exactly $n$ times around the nucleus  in one laser cycle, and we refer them as 1:$n$ resonances). These resonances give birth to periodic orbits, among which at least one, denoted $\mathcal{O}_{n}$, is unstable (hyperbolic). Other periodic orbits in the vicinity of the bound region merely influence the fine details of the chaotic structure.

\begin{figure}
	\centering
		\includegraphics[width = \linewidth]{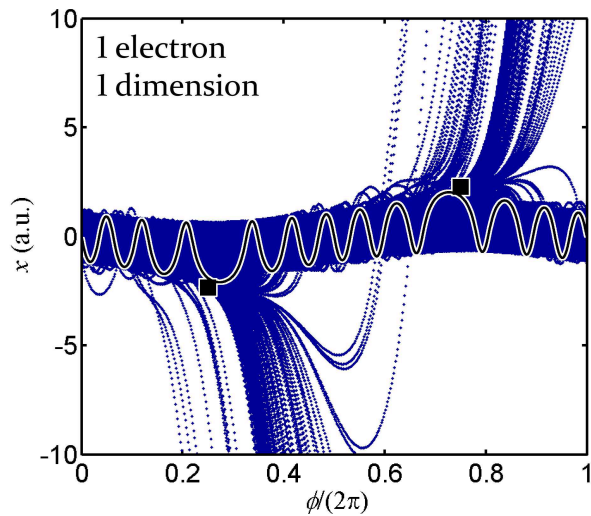}
	\caption{\label{fig:Manifold}
	Projection of the unstable manifold of the periodic orbit $\mathcal{O}_{12}$ of Hamiltonian~(\ref{eq:He+}) for $I=3\times 10^{15}\ {\rm W}\cdot{\rm cm}^{-2}$ and $780$~nm in the $(x,\phi)$ plane. The black curve is a projection of the periodic orbit in the plane $(x,\phi)$, and the black squares indicate the position of the saddle point at the maximum of the field.}
\end{figure}

Because a periodic orbit corresponds to a recurrent motion, it does not lead to ionization. However, neighboring trajectories may do so after some time, particularly if the periodic orbit is weakly unstable, as it is the case for selectes $\mathcal{O}_{n}$s. The pathways by which the electrons approach or leave the core are quantified by the so-called stable and unstable manifolds~\cite{IntroAppliedNonlinDynSystChaos} around an unstable periodic orbit. In Fig.~\ref{fig:PSOS_RESI1} (lower panel) we draw the stable and unstable manifolds of $\mathcal{O}_{n}$ (light orange and dark blue points respectively) for $n=12$. Note the strong similarity between the unstable manifold and the Poincar\'e sections of RESI trajectories for the two-electron Hamiltonian~(\ref{eq:Ham2e}) (upper panel). This similarity confirms the key role played by this one-electron unstable manifold in the RESI process. In addition, we see that the stable and unstable manifolds intersect an infinite number of times, a characteristic feature of a chaotic dynamics~\cite{Mitc09}. The overlap between the stable and unstable manifolds of $\mathcal{O}_{n}$ forms a ``sticky'' region~\cite{IntroAppliedNonlinDynSystChaos} that traps trajectories for some time before ionizing. To better understand how RESI electrons find their way to ionization, we display a projection of the unstable manifold of $\mathcal{O}_{n}$ in the position--phase of the laser plane $\left(x,\phi=\omega t+\phi_{0}\right)$, together with a projection of $\mathcal{O}_{n}$ in Fig.~\ref{fig:Manifold}. Two main branches depart from the central region near $x=0$ when the laser phase is $\phi=\pi/2$ and $\phi=3\pi/2$, i.e., near the extrema of the electric field, showing that RESI takes place approximately at the extrema of the electric field~\cite{Mosh00,Rude04}.

\begin{figure}
	\centering
		\includegraphics[width = \linewidth]{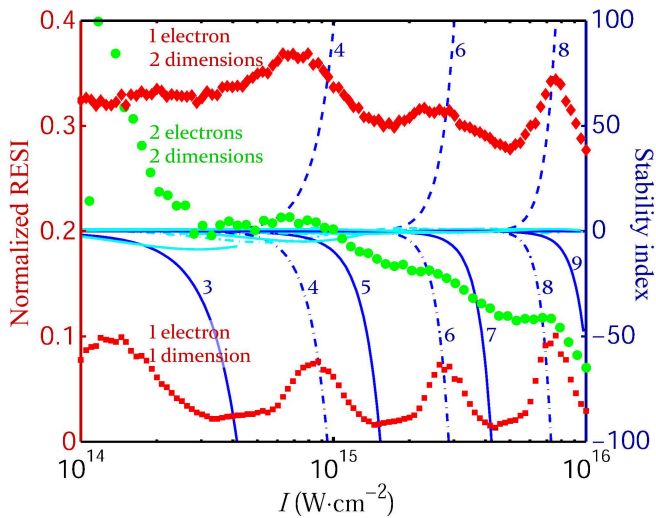}
	\caption{\label{fig:RESI_to_DI_ratio}
	Normalized RESI yields for 460~nm wavelength laser (markers, left hand $y$-axis). Red squares (diamonds) label one (two) dimensional one-electron simulations (normalized to the number of non-ionized trajectories). Green dots label two-dimensional two-electron simulations (normalized to the number of double ionization). Curves label the linear stability index~\cite{IntroAppliedNonlinDynSystChaos} of the resonant periodic orbits $\mathcal{O}_{n}$ of Hamiltonian~(\ref{eq:He+}) (curves, right hand y-axis) as functions of the laser intensity~$I$. Continuous (broken) curves refer to odd (even) 1:$n$ resonances. As intensity increases, the order of the resonance goes from $n=3$ to $n=9$ as indicated on the curves.}
\end{figure}

The organizing role of resonant periodic orbits can also be found in the relative proportion of RESI to double ionization yields: In Fig.~\ref{fig:RESI_to_DI_ratio}, we compute the RESI yields as the intensity is varied. It shows oscillations which are correlated with the intensity range where the 1:$n$ resonance regulates the delayed ionization. The bound region (pink area in Fig.~\ref{fig:PSOS_RESI1}) shrinks with increasing laser intensity, and higher-order resonant periodic orbits are drawn to the unbound region. It means that as $\mathcal{O}_{n}$ becomes too unstable, the next resonance (associated with periodic orbit $\mathcal{O}_{n+1}$) is at play. Note the correlation between the oscillations and the stability index~\cite{IntroAppliedNonlinDynSystChaos} of the resonant periodic orbits identified from the one-electron model~(\ref{eq:He+}). The stability index measures the typical time this orbit is expected to influence the neighboring dynamics: The larger the stability index, the sooner a neighboring trajectory diverges from it. Recent experimental results~\cite{Pfei11}, with elliptic polarization, have revealed oscillations in the parallel to anti-parallel double ionization yields analogous to the oscillations observed in the RESI yields of Fig.~\ref{fig:RESI_to_DI_ratio}.

Adding a dimension to the one-dimensional calculations provides more scope for the electron dynamics and makes the dynamics more complex. In particular, due to the extra dimension, the dynamics close to the nucleus region becomes partially chaotic both for one- and two-electron Hamiltonians~(\ref{eq:He+}) and~(\ref{eq:Ham2e}) as it is already seen in Fig.~\ref{fig:RESI_mechanism}. It becomes harder to define a bound and unbound regions for the inner electron. More importantly, the aforementioned 1:$n$ resonances generally enhance (linear) instability in the transverse direction while not affecting the organization of the dynamics in the (polarization) symmetry subspace: At a given intensity more than one resonance drives the RESI dynamics in the sense that their unstable directions  participate to the enhancement of delayed ionization. A direct consequence is the smoothing of the oscillations in Fig.~\ref{fig:RESI_to_DI_ratio} for $460$~nm wavelength. Increasing the wavelength to $780$~nm gives birth to more resonances with a more dense tangle of unstable manifolds so that the oscillations are completely washed out in two-dimensional models whether they are one- or two-electron models, as observed from our classical calculations.

In Fig.~\ref{fig:RESI_to_DI_ratio} we notice that the RESI yields are higher for a two-dimensional model than for one dimension. We attribute these higher yields to the aforementioned chaotic dynamics coming from an increase of instability associated with additional resonances in the full-dimensional models. It underscores once more the pivotal role played by the inner electron dynamics [Hamiltonian~(\ref{eq:He+}), i.e., without recollision] in these delayed ionizations. 

The authors acknowledge useful discussions with C. Cirelli, N. Johnson, R.~R. Jones, U. Keller, M. Kling, and A. Pfeiffer. C.C. and F.M. acknowledge financial support from the CNRS. F.M. acknowledges financial support from the Fulbright program. This work is partially funded by NSF.



\end{document}